\documentclass[%
 reprint,
amsmath,amssymb,
aps,
pra,
floatfix,
]{revtex4-2}
\usepackage{placeins} 
\usepackage{graphicx}
\usepackage{dcolumn}
\usepackage{bm}
\usepackage{physics}
\usepackage[T1]{fontenc}

\begin{document}

\preprint{APS/123-QED}

\title{Entanglement dynamics of Multi-Level Atoms embedded in Photonic Crystals: Leveraging Resonant Dipole-Dipole Interactions and Quantum Interference}


\author{Nancy Ghangas}
 \email{nancy.19phz0006@iitrpr.ac.in}
\author{Shubhrangshu Dasgupta}
\affiliation{%
Department of Physics, Indian Institute of Technology Ropar, Rupnagar, Punjab, India\\
}%




\date{\today}

\begin{abstract}
 We present a comprehensive investigation of entanglement dynamics in multi-level V-type atomic systems embedded within photonic crystals. We mainly focus on the synergistic roles of resonant dipole-dipole interactions and quantum interference through analytical modeling and numerical simulations using the Schrodinger equation. Key findings reveal that resonant interaction dominates when the interatomic distance is comparable to the localization length of photon-atom bound states lying in the bandgap region. For atoms with anti-parallel dipole orientations, both initially entangled and separable states exhibit robust entanglement preservation due to strong collective interactions. Conversely, when dipoles are oriented orthogonally, initially entangled states exhibit unique oscillatory patterns in their entanglement dynamics. This effect arises from the formation of dark states due to destructive interference within the structured photonic environment, with resonant dipole-dipole interactions sustaining non-Markovian dynamics. We further demonstrate that positioning the atomic excited states deeper within the photonic bandgap accelerates the decay of entanglement oscillations due to the exponential suppression of resonant energy exchange mediated by evanescent modes. Our analysis establishes resonant dipole-dipole interactions and quantum interference as potential tools for tailoring entanglement dynamics, paving the way for controlled quantum coherence in photonic crystal platforms.
\end{abstract}

\maketitle


\section{\label{sec:level1}Introduction}

Cavity quantum electrodynamics (QED) plays a central role in exploring and engineering light-matter interactions. In last few decades, cavity QED systems have been extensively used in quantum computing and information processing and exploiting nonclassical resources in such tasks. In 1987, 
Yablonovitch had shown that a photonic crystal (PC) can serve as a cavity and 
the radiative decay rate of an emitter gets affected by the electromagnetic environment inside such a cavity \cite{PhysRevLett.58.2059}.
A strong coupling of an atom with a PC can lead to photon-atom bound state (leading to splitting of excited states) near band-edge, an effect analogous to polaritons in usual cavity QED systems 
\cite{PhysRevLett.64.2418}. In this scenario, spontaneous emission in the band gap is inhibited and the hybrid dressed states in the absence of any external field is stabilizd, along with a spatial localization of field around the emitter \cite{litinskaya2016broadband,liu2017quantum}.
Expanding upon the pioneering contributions of Sajeev John \cite{PhysRevB.43.12772}, Yang and Zhu \cite{PhysRevA.61.043809,PhysRevA.64.013819} built upon this framework by analyzing how quantum interference influences the spontaneous emission dynamics in both isotropic and anisotropic 1D PC environments. These contributions have shaped the motivation behind our investigation of multi-level atomic systems in structured photonic environments to achieve tunable quantum correlations.

In cavity QED, symmetric/antisymmetric atomic superpositions can form robust entangled states, yet environmental noise may trigger entanglement sudden death well before the spontaneous emission kicks in \cite{PhysRevLett.93.140404,article70,man2015cavity}. Entanglement can undergo sudden death between qubits embedded in separate cavities \cite{article,doi:10.1126/science.1142654,doi:10.1126/science.1167343} or emerge suddenly (sudden birth of entanglement) due to collective dissipative dynamics \cite{article100}. In multi-qubit systems, entanglement evolution is non-trivially complex and cannot be inferred from single-qubit decoherence \cite{PhysRevLett.93.140404,Yönaç_2007}.
Quantum entanglement can be generated and controlled across various photonic platforms, for instance PC cavity systems use driving fields \cite{article20,Kim:11} or virtual photons \cite{PhysRevLett.87.037902}, while solid-state systems employ electrical tuning \cite{PhysRevB.82.075305} or external-field-driven magnon baths \cite{PhysRevResearch.4.023221}. Advanced quantum control leverages off-resonant cavity QED gates and bichromatic lasers to achieve high-fidelity operations and stabilized entanglement \cite{Akbari_2017,Antón_2021,yan2025cavity}.

PC platforms support integration of experimentally demonstrated quantum emitters such as diamond nitrogen vacancy (NV) centers \cite{katsumi2025recent,PhysRevResearch.4.023221,yurgens2024cavity}, trapped neutral atoms \cite{PhysRevLett.115.063601,doi:10.1073/pnas.1603788113,kim2019trapping,PhysRevA.78.033429,doi:10.1126/science.abi9917,PhysRevLett.124.063602}, semiconductor quantum dots \cite{calic2017deterministic,PhysRevLett.101.113903,PhysRevB.79.205416,PhysRevX.12.021042}, and Rydberg atom systems \cite{PhysRevA.102.022816,PhysRevResearch.5.043135,epple2014rydberg}. Numerous classic studies have also demonstrated the use of PCs with atomic ensembles for practical applications, including quantum information processing and novel photonic devices \cite{GHANGAS2025131876,PhysRevX.4.031022}.
However, PC cavities excel in quantum applications by preserving spin squeezing in qubit ensembles \cite{Zhong_2016}, generating fidelities exceeding $80\%$ for entangled photon pairs from quantum dots \cite{PhysRevB.79.205416}, and enabling quantum optical switches for integrated networks \cite{article50,wang2016relationship}. They also support Rydberg giant-atom platforms \cite{PhysRevResearch.5.043135} and multi-qubit entanglement dynamics in GHZ states across 1D/3D systems, surpassing traditional nanocavities in robustness and functionality \cite{PhysRevA.80.022314,PhysRevA.85.014301}.

Investigating the collective quantum dynamics of multiple atoms embedded within PCs elucidates the interplay of resonant dipole-dipole interactions (RDDI), quantum interference (multi-level atoms), and bandgap-mediated coupling in structured photonic environments.
Photonic tunneling creates impurity bands in the bandgap, where two excited atoms exhibit distance-dependent RDDI showcasing inhibition, enhancement, or oscillations with exponential decay as reported in \cite{PhysRevA.42.2915,PhysRevB.43.12772}.
A recent study has investigated that RDDI decays as the inverse of interatomic distance due to suppressed resonant photon exchange in a 1D photonic band gap (PBG), mediated instead by low-frequency evanescent modes \cite{article60,PhysRevA.42.2915} in contrast with \cite{PhysRevLett.61.2269,PhysRevLett.77.406,PhysRevLett.77.407}. 
PBG engineering enables subnatural-linewidth spectral splitting \cite{PhysRevA.50.1764} and band-edge-localized superradiance in atomic ensembles \cite{PhysRevLett.74.3419}. Disordered dipoles form spin-glass phases with optical bistability \cite{PhysRevLett.76.1320}, while weak pumping induces collective inversion \cite{PhysRevLett.78.1888} and selective emission control in multi-level atoms \cite{PhysRevLett.79.5238}.
RDDI among two-level atoms within PBG environments enhances resonant interatomic forces near the photonic band edge, governing entanglement dynamics with/without external fields \cite{WANG20115323,WU201574,PhysRevA.89.062117}.

In this work, we demonstrate the collective quantum dynamics of two multi-level V-type atoms embedded in a common PC while prior studies have explored atom-photon entanglement for single V-type atom embedded within PCs \cite{PhysRevA.64.013819,Li:15,SAHRAI2017116,vafafard2017vacuuminducedcoherencecavity}. We uniquely investigate the combined impact of RDDI and quantum interference in PCs, a regime unexamined in previous cavity or free-space studies \cite{article20,PhysRevA.63.043805,PhysRevA.74.032313,PhysRevA.81.052341,PhysRevA.52.2835}.
To our knowledge, this study presents the first comprehensive analytical and numerical report of atom-atom quantum correlations for two multi-level atoms embedded within a PC system unlocking unprecedented command over entanglement evolution. 
With the consideration of entangled and unentangled initial state preparations of the atomic pair, it reveals two pivotal findings: ($1$) Enhanced RDDI strength significantly prolongs entanglement preservation across antiparallel dipole transition configurations demonstrating the universal role of RDDI in countering decoherence, ($2$) For orthogonal dipole orientations in the case of the entangled initial state, entanglement dynamics display periodic oscillations, a hallmark of non-Markovian behavior. Further, the frequency and amplitude of entanglement oscillations are critically governed by the detuning of atomic excited-state energies relative to the photonic band-edge frequency and initial quantum state preparation. In contrast, atomic system initialized in separable quantum unentangled states exhibit accelerated entanglement decay under orthogonal dipole alignments. Consequently, the synergy of quantum interference and RDDI leads to the preservation of quantum coherence and quantum correlations through their cooperative dynamics even in deep gap regions.
 
This paper is organized as follows. The analytical model to analyze atom-atom entanglement dynamics is presented in section~\ref{sec:2}. In section~\ref{sec:3}, numerical and analytical analysis of the combined effect of RDDI and quantum interference in the preservation of entanglement dynamics is given. We conclude the paper in section~\ref{sec:4}.

\section{Theoretical formalism}
\label{sec:2}
\begin{figure}[!htbp]
\centering
\includegraphics[width=\linewidth]{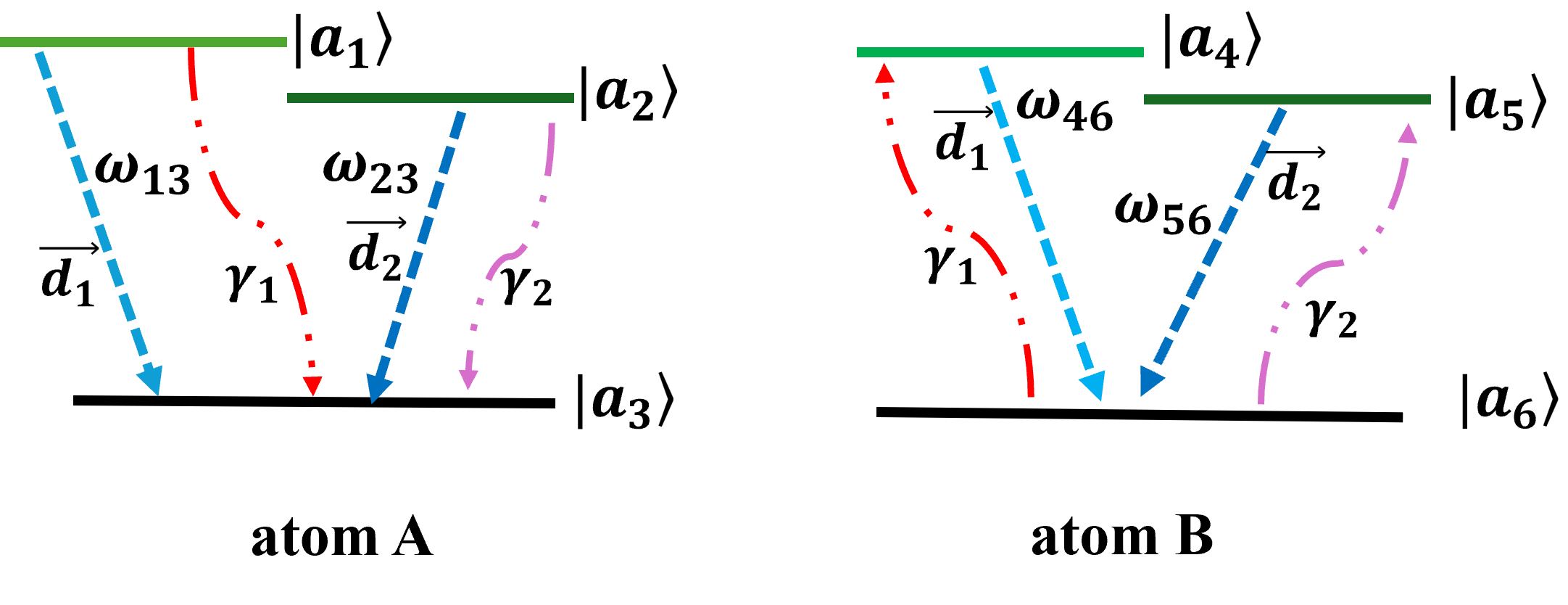}
\caption{Schematic of two V-type three-level atoms.}
\label{fig.1}
\end{figure}
We consider a system of two three-level V-type atoms A and B (as shown in Fig. \ref{fig.1}) embedded in an isotropic 1D PC. The transition frequency between the ground state $\ket{a_3}$ ($\ket{a_6}$) and the two upper levels, $\ket{a_1}$ and $\ket{a_2}$ ($\ket{a_4}$and $\ket{a_5}$) of atom A (atom B) are given by $\omega_{13}$ ($\omega_{46}$) and $\omega_{23}$ ($\omega_{56}$), respectively. We assume that the two atoms are identical and therefore, $\omega_{13}=\omega_{46}$ and $\omega_{23}=\omega_{56}$.  These transition frequencies are assumed to be lying near the photonic band edge. Considering dipole approximation and rotating wave approximation, the total Hamiltonian of the system can be written as, ($\hbar =1$)
\begin{equation}
   \hat{H}_{total} = \hat{H}_{atom} + \hat{H}_{field} + \hat{H}_{int} + \hat{H}_{DDI},
\end{equation}
where,
\begin{eqnarray*}
       \hat{H}_{atom} = [\omega_{13} \left(\ket{a_1 a_6 }\bra{a_1 a_6 }+ \ket{a_3 a_4}\bra{a_3 a_4 } \right) \\
        + \omega_{23} (\ket{a_2 a_6 }\bra{a_2 a_6 } + \ket{a_3 a_5 }\bra{a_3 a_5 } ) ]\otimes \mathbf{I} \;,
\end{eqnarray*}
and
\begin{eqnarray*}
    \hat{H}_{field} =\sum_k \omega_k a_k^\dagger a_k \;,
\end{eqnarray*}
are the unperturbed Hamiltonian for the two atoms and the field, respectively.  Here, $\mathbf{I}$ is the identity operator in the Hilbert space of the radiation modes, and $a_k$ and $a_k^\dagger$ are the annihilation and creation operators, respectively, for the $k^{th}$ radiation mode with frequency $\omega_k$ in the PC. The interaction Hamiltonian $\hat{H}_{int}$ is given by
\begin{eqnarray*}
   \hat{H}_{int} =\sum_k\left[ {g_k}^{(1)} ( \ket{a_3 a_6}\bra{a_1 a_6 }+ \ket{a_3 a_6 }\bra{a_3 a_4 })\ket{1_k}\bra{0_k} \right. \\
   \left. + {g_k}^{(2)} ( \ket{a_3 a_6}\bra{a_2 a_6 }+ \ket{a_3 a_6 }\bra{a_3 a_5 })\ket{1_k}\bra{0_k}+ h.c.\right] \;,
\end{eqnarray*}
where $g_k^{(1)}$ ($g_k^{(2)}$)
are the coupling constants between the $k^{th}$ electromagnetic mode and the atomic transitions $\ket{a_1}\leftrightarrow \ket{a_3}$ and $\ket{a_4}\leftrightarrow \ket{a_6}$  ($\ket{a_2}\leftrightarrow \ket{a_3}$ and $\ket{a_5}\leftrightarrow \ket{a_6}$). 
These transitions occur in one of the atoms, while the other atom remains in its ground state. 
The dipole-dipole interaction Hamiltonian $\hat{H}_{DDI}$ is given by
\begin{equation*}
   \hat{H}_{DDI} = (\gamma_1 \ket{a_1 a_6 }\bra{a_3 a_4 } 
   + \gamma_2 \ket{a_2 a_6 }\bra{a_3 a_5 } + h.c.)\otimes \mathbf{I} \;.
\end{equation*}
Here, $\gamma_i$ ($i\in 1,2$) is the dipole-dipole coupling strength, for the energy transfer from one atom to the other without any mediation of the electromagnetic modes.


In an isotropic PC, the dispersion relation near the right band edge can be expressed in the form $\omega_k=\omega_c + A(k-k_0)^2$, where $A=\omega_c/k_0^2$ is a constant coefficient, $\omega_c$ is the cutoff frequency of band edge, and $k_0=\omega_c/c$, represents branch cut singularity in complex $k$-plane \cite{PhysRevA.50.1764,PhysRevB.43.12772}. It is to be noted that our model is restricted to the single‐photon manifold and exclusively treats the zero‐phonon transition; contributions from any phonon sidebands have been neglected. Therefore, the state vector of the system at time $t$ can be written as $(\iota = \sqrt{-1})$
\begin{eqnarray}
|\psi(t)\rangle &=& A_1(t) e^{\iota \omega_{13}t} \ket{a_1 a_6 {0}_k} + A_2(t) e^{\iota \omega_{23}t} \ket{a_2 a_6 {0}_k} \nonumber\\ 
&&+ A_3(t) e^{\iota \omega_{13}t} \ket{a_3 a_4 {0}_k} + A_4(t) e^{\iota \omega_{23}t} \ket{a_3 a_5 {0}_k} \nonumber\\
&&+ \sum_kB_k(t) e^{\iota \omega_{k}t} \ket{a_3 a_6 {1}_k} \;,
\end{eqnarray}
where the state vectors $\ket{a_i a_j 1_k}$ describe atom A in $\ket{a_i}$ state, atom B in $\ket{a_j}$, and radiation mode in $\ket{1_k}$ state (i.e., a state with one photon in the $k$th mode and no photon in all the other modes). We assume that atom+field state is initially at the zero-photon subspace, such that $B_k(0)=0$, $\forall k$. 
Using the Schrodinger equation, $\iota\hbar\frac{\partial}{\partial t}\ket{\psi(t)} = \hat{H}_{total}\ket{\psi(t)}$, we get the following set of first-order coupled differential equations for the probability amplitudes $A_i(t)$ and $B_k(t)$:
\begin{eqnarray} 
\iota\frac{\partial A_1(t)}{\partial t} &=& \gamma_1 A_3(t)  + \sum_k g_k^{(1)} B_k(t)  e^{-\iota (\omega_{k}-\omega_{13})t} \label{eq:2}\;,\\
   \iota\frac{\partial A_2(t)}{\partial t} &=& \gamma_2 A_4(t)  + \sum_k g_k^{(2)} B_k(t)  e^{-\iota (\omega_{k}-\omega_{23})t} \label{eq:3}\;,\\
    \iota\frac{\partial A_3(t)}{\partial t}& =& \gamma_1 A_1(t) + \sum_k g_k^{(1)} B_k(t)  e^{-\iota (\omega_{k}-\omega_{13})t} \label{eq:4}\;,\\
    \iota\frac{\partial A_4(t)}{\partial t} &=& \gamma_2 A_2(t)  + \sum_kg_k^{(2)} B_k(t)  e^{-\iota (\omega_{k}-\omega_{23})t} \label{eq:5}\;,\\
\iota\frac{\partial B_k(t)}{\partial t} &= &g_k^{(1)}\left[A_1(t)  + A_3(t)\right] e^{-\iota (\omega_{13} -\omega_{k})t}\nonumber\\
&& +g_k^{(2)}\left[A_2(t)  + A_4(t)\right] e^{-\iota (\omega_{23}-\omega_{k})t} \label{eq:6}\;.
\end{eqnarray}

We have analytically solved these equations for $A_i(t)$ for all $i$ by using Laplace transformations and have included the results in the Appendix. From these expressions (\ref{A.10}-\ref{A.13}), we see that the probability amplitudes are dependent on the angle $\eta$ through the cross-damping term $\Gamma_{ij}$ [see Eq.(\ref{gammaij})]. This angle between two transition dipole moments $\vec{d}_{13}$ ($\vec{d}_{46}$) and $\vec{d}_{23}$ ($\vec{d}_{56}$) of the atom A (atom B) leads to the so-called vacuum-induced coherence between the energy levels $\ket{i}$ and $\ket{j}$.
Using these solutions (\ref{A.10}-\ref{A.13}), we next obtain the time-dependent density matrix, $\rho(t) =\ket{\psi(t)}\bra{\psi(t)}$ of the two atoms and the radiation modes. 

To study the dynamics of the atom-atom entanglement, we obtain the reduced density matrix of the two atoms and calculate the logarithmic negativity $E_N$.
In our analysis of V-type atoms in PCs, this $E_N$ quantifies the entanglement mediated by RDDI and quantum interference. Its insensitivity to specific state purity makes it ideal for analyzing disordered or lossy environments, where traditional measures fail. For instance, in bandgap-engineered PCs, $E_N$ captures entanglement dynamics near photonic bound states, in the presence of structural disorder or non-Markovian decay \cite{PhysRevA.65.032314,article400}. Alternative measure of entanglement, namely, the von Neumann entropy of the atom-field system in a PC has also been explored in \cite{Li:15}. However, the negativity poses as an appropriate measure for quantifying entanglement in two-qutrit systems \cite{Li:15}.

For a bipartite quantum state $\rho_{AB}$, the logarithmic negativity $E_N$ is defined as $E_N(\rho_{AB})=\log_2||\rho_{AB}^\Gamma||_1$, where $\rho_{AB}^\Gamma$ is the partial transpose of $\rho_{AB}$ with respect to the subsystem B (radiation field, in the present case), $||.||_1$ denotes the trace norm (sum of singular values) equivalent to absolute sum of negative eigenvalues $\lambda_i$ of $\rho_{AB}^\Gamma$ such that
$E_N(\rho_{AB})=\log_2 \left(1+2\sum_{\lambda_i <0}|\lambda_i|\right)$.

\section{Results and Discussion}\label{sec:3}
In this section, we will discuss how the logarithmic negativity between two V-type atoms embedded inside a PC evolves with time. We focus on two distinct configurations: anti-parallel and orthogonal orientations of the atomic dipole transitions with each other. For each configuration, we study the entanglement dynamics of both initially unentangled and initially entangled atomic states to understand how initial correlations affect the system’s evolution. Furthermore, we demonstrate how varying the transition frequencies of the atoms relative to the photonic band edge profoundly influences the dynamics of entanglement.

We first consider the case of anti-parallel alignment of the atomic dipoles, i.e., $\eta=\pi$. The corresponding dynamics of $E_N$ is displayed in Fig. \ref{fig.2} and Fig. \ref{fig.4} for different strengths of RDDI.
We observe that the entanglement is retained for longer times, for larger RDDI. The RDDI creates a coherent coupling even in the absence of real photons. In the PBG environment near the band edge, spontaneous emission is inhibited due to which the RDDI becomes dominant. The atoms cannot emit into the PC continuum and the RDDI prepares them into entangled states, which are immune to the decay.
It is to emphasize that such behavior has already been predicted for atom-atom separations much larger than the optical wavelength for different ($\Sigma$ and $\Pi$) states in PCs \cite{PhysRevB.43.12772}. In our case, the interatomic distances can be of the order of micrometers. The coupling rate $\gamma_1=\gamma_2=10\beta$ [$\beta$, having a dimension of frequency, is defined in Eq.(\ref{beta}), and is used as the normalizing parameter throughout this paper]  corresponds to an effective interaction range of the order of the localization length, which enables resonant energy exchange and prolonged entanglement.

For $\eta=\pi$ configuration, we observe that the atomic excited states split into four localized dressed states. The functions $G_i(x)'s$ [see Eq.(\ref{A.5})] exhibit one pure imaginary pole lying in the band pass region, while the functions $H_i(x)'s$ [see Eq.(\ref{A.6})] yield three pure imaginary roots in the band gap region [see Fig. \ref{fig.3}]. This means that the three localized states lying in the band gap region sustain photon-atom bound states while one leaky state lying out of the band gap region with finite density of states (DOS) allows residual interactions.

For orthogonal alignment of atomic dipoles, $\eta=\pi/2$, we observe that the functions $G_i(x)'s$ exhibit one pure imaginary root lying in the band pass region and the functions $H_i(x)'s$ yield two pure imaginary roots in the band‐gap region. We note that the probability amplitudes $A_{2,4}(t)$ contribute trivially to the overall correlations with their magnitude remaining zero, while $A_{1,3}(t)$ contribute to the system with an amplitude that decays exponentially over time. As a result, the time-dependent density matrix exhibits significantly reduced coherence terms compared to $\eta=\pi$ case with distinct interference effects for different choices of initial states. 

We note that the coherence between two V-type atoms in free space can be achieved even when their dipole moments are orthogonal \cite{PhysRevA.63.043805}. The effect of initial state in different atomic configurations has been further studied in \cite{DERKACZ20087117} in this context. 
However, these results pertain to atomic separations on the order of the optical wavelength. In contrast, we consider an atom-atom separation comparable to the localization length, where the PBG environment modifies interference effects by confining interactions to atom–photon bound states.

In the subsequent sections, we investigate how different choices of initial states affect the entanglement dynamics.

\subsection{Unentangled state as initial state}

\begin{figure*}[!htbp]
\centering
\includegraphics[width=0.9\linewidth]{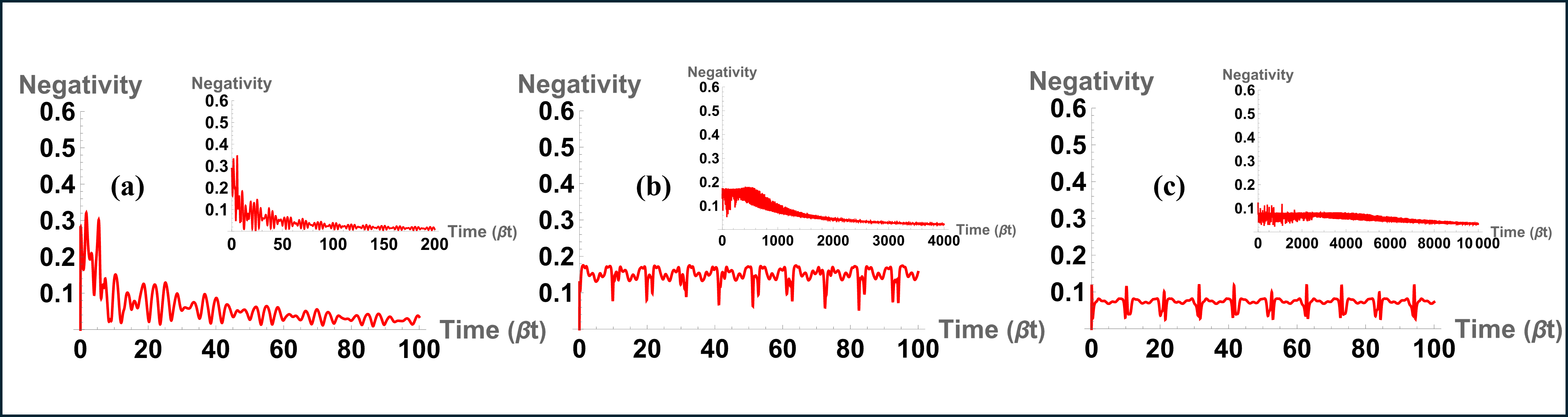}
\caption{Negativity dynamics as a function of $\beta t$, for unentangled initial state as $\psi(0)=\ket{a_1,a_6}$, for position of dipole transitions to be anti-parallel, $\eta=\pi$ such that $\omega_{1c}=0.6\beta$, $\omega_{2c}=0.2\beta$, and $\omega_{12}=0.4\beta$ for RDDI strengths, (a) $\gamma_1=\gamma_2=1.5\beta$, (b) $\gamma_1=\gamma_2=6\beta$, (c) $\gamma_1=\gamma_2=10\beta$.}
\label{fig.2}
\end{figure*}

We consider the case of an unentangled state as the initial state, i.e., $\ket{\psi(0)}= \ket{a_1,a_6}\otimes\prod_k\ket{0_k}$. We choose the $\eta=\pi$ configuration, and $\omega_{1c}=0.6\beta$ and $\omega_{2c}=0.2\beta$, which represent the difference between the atomic transition frequency and the band edge.
We observe that the atoms get entangled as soon as $t>0$ [see Fig. \ref{fig.2}]. For the  RDDI strengths  $\gamma_1=\gamma_2=1.5\beta$, the oscillations in negativity dynamics decay quickly at a time-scale $\beta t=60$ [see Fig. \ref{fig.2}(a)]. While for the stronger RDDI, e.g., $\gamma_1=\gamma_2=6\beta$, the entanglement is sustained till $\beta t=1000$ [see Fig. \ref{fig.2}(b)]. On further increasing RDDI strength to $\gamma_1=\gamma_2=10\beta$, the entanglement is maintained for longer times (up to $\beta t=5000$ [see Fig. \ref{fig.2}(c)]).

\begin{figure*}[!htbp]
\centering
\includegraphics[width=0.75\linewidth]{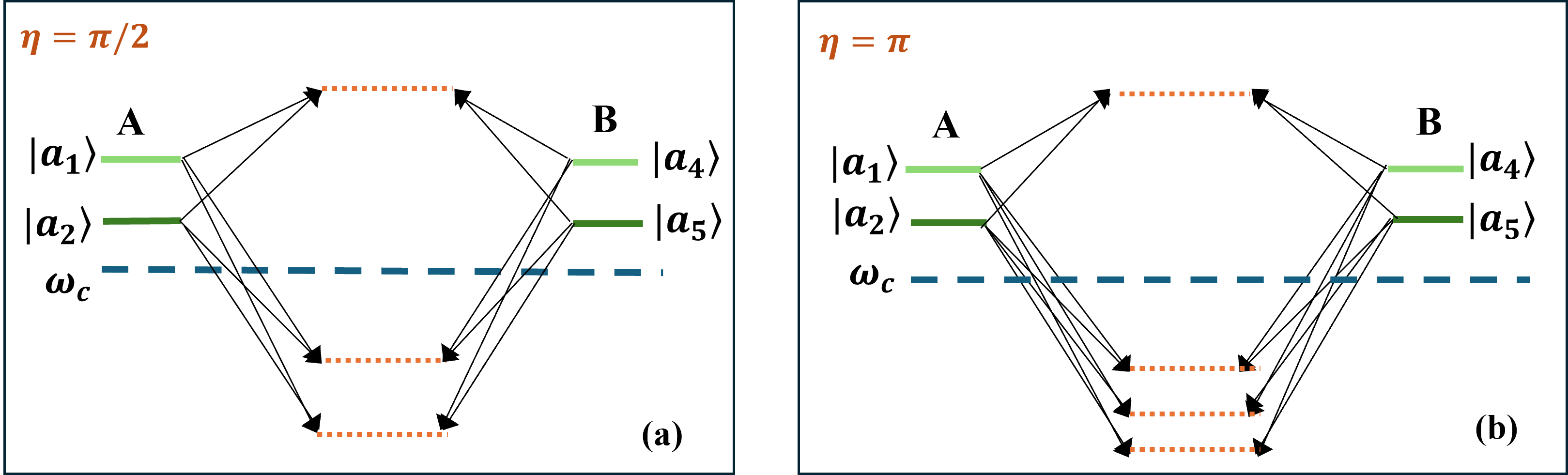}
\caption{Schematic of atomic splitting for unentangled state as initial state for the position of $\omega_{1c}=0.6\beta$, $\omega_{2c}=0.2\beta$ relative to $\omega_c$, where $\omega_c$ lie below $\ket{a_1}(\ket{a_4})$ and $\ket{a_2}(\ket{a_5})$, and $\omega_{12}=0.4\beta$, when dipole transitions of excited levels $\ket{a_1}(\ket{a_4})$ and $\ket{a_2}(\ket{a_5})$ are, (a) orthogonal, $\eta=\pi/2$, (b) anti-parallel, $\eta=\pi$.}
\label{fig.3}
\end{figure*}

These results can be explained in terms of the dressed states. Analytically, we observe that the position of dressed states in the band gap occurs at $-3.4\beta\iota$, $-6\beta\iota$, and $-5.6\beta\iota$ while the dressed state at $6\beta\iota$ lies in the band pass region for $\gamma_1=\gamma_2=6\beta$ [see Fig. \ref{fig.3}(b)]. It is to be noted that the number and characteristics of dressed states can vary for different RDDI strength choices.
Therefore, the entanglement is sustained for longer times at larger RDDI strength, as the dressed states shift away from the band edge.  
As a result, it creates population imbalance such that there is less overlap in superposition states, and not all the pathways remain in phase for constructive interference. Hence, it leads to less sharp entanglement buildup in the PC system.
However, several off-diagonal non-decaying coherence terms in the density operator maintain phase-sensitive interference, counteracting environmental dephasing to preserve entanglement lifetimes.

Further, for the case of orthogonal dipole moments, the dressed states lie at $6\beta\iota,-6\beta\iota,-5.6\beta\iota$ as shown in Fig. \ref{fig.3}(a). Since the exchange of population also involves non-band gap mode, the coupling of this mode leads to the fast exponential decay of coherence terms in the time-dependent density matrix. Therefore, due to the lack of initial quantum correlations, there is the induction of destructive interference between vacuum-mediated pathways. In addition, these non-band gap modes experience residual photonic DOS which accelerates decoherence and suppresses coherent energy exchange (that could happen via RDDI). Hence, negativity decays faster and cannot maintain population oscillations for extended periods. 

\subsection{Entangled bright state as initial state}
\begin{figure*}[!htbp]
\centering
\includegraphics[width=0.9\linewidth]{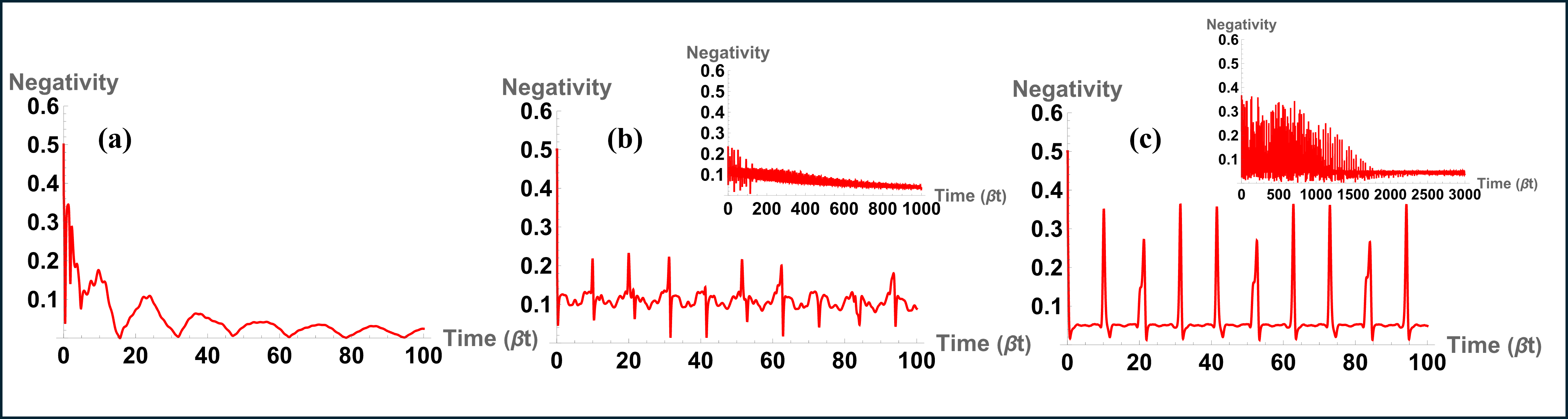}
\caption{Negativity dynamics as a function of $\beta t$, for position of dipoles, $\omega_{1c}=-0.6\beta$, $\omega_{2c}=-1\beta$, and $\omega_{12}=0.4\beta$ for anti-parallel, $\eta=\pi$ dipole alignments for different RDDI strengths, (a) $\gamma_1=\gamma_2=1.5\beta$, (b) $\gamma_1=\gamma_2=6\beta$, (c) $\gamma_1=\gamma_2=10\beta$.}
\label{fig.4}
\end{figure*}

Now, we consider the case when atoms are initially prepared in an entangled bright state, i.e., $\ket{\psi(0)}=\frac{1}{\sqrt 2}(\ket{a_1,a_6}+\ket{a_3,a_4})\otimes\prod_k\ket{0_k}$. The excited states ($\ket{a_1} (\ket{a_4})$, $\ket{a_2} (\ket{a_5})$) lie within the PBG, with $\omega_{1c}=-0.6\beta$ and $\omega_{2c}=-1\beta$. In this configuration, the atom A prepared in an excited state can emit a photon via its transition from $\ket{a_1}$ to $\ket{a_3}$ so that the bound photon supported by the long localization length inside the gap tunnels through the crystal and is reabsorbed by the atom B, promoting it from $\ket{a_6}$ to $\ket{a_4}$. Because the density of propagating modes vanishes in the gap, this exchange is mediated entirely by evanescent, non-radiative channels.
We note that as RDDI strength is increased from $\gamma_1=\gamma_2=1.5\beta$ to $10\beta$, it extends the time over which their logarithmic negativity survives till $\beta t\le50$ and $\beta t=2000$, respectively, as shown in Fig. \ref{fig.4}(a)-(c), thus preserving entanglement for longer intervals.

Analytically, we observe that poles of $H_i(x)'s$ lie at $-4.6\beta\iota$, $-6\beta\iota$, $-5.6\beta\iota$ in the band gap region and that of $G_i(x)'s$ lies at $6\beta\iota$ which is in the band pass region [see Fig. \ref{fig.6}(b)]. We also note that the contribution from the dressed states lying within the pass band is suppressed and the exchange of population only among band gap modes enhances the $E_N$. This results in long-lived entanglement compared to the case of unentangled initial states as discussed before. 
In addition, the stronger RDDI ($\gamma_1=\gamma_2=6\beta,10\beta$) [see Fig. \ref{fig.4}(b),(c)] amplifies this interference, sustaining oscillations and entanglement over longer times, whereas the weaker RDDI ($\gamma_1=\gamma_2=1.5\beta$) allows rapid decay and loss of entanglement as shown in Fig. \ref{fig.4}(a). 

\begin{figure*}[!htbp]
\centering
\includegraphics[width=0.9\linewidth]{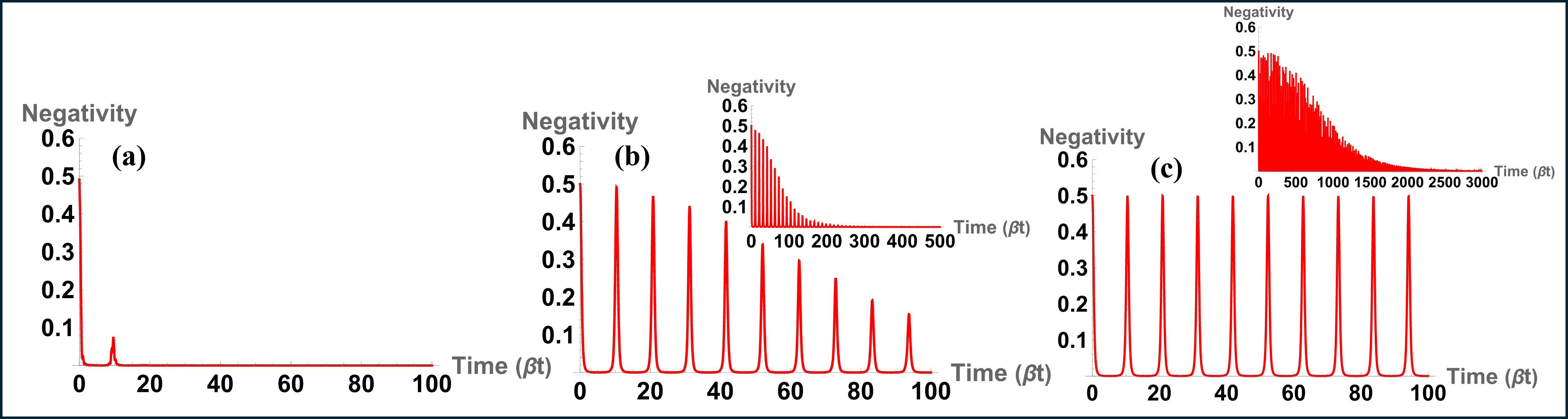}
\caption{Negativity dynamics as a function of $\beta t$, for position of dipoles, $\omega_{1c}=-0.6\beta$, $\omega_{2c}=-1\beta$, and $\omega_{12}=0.4\beta$ for orthogonal dipole transitions, $\eta=\pi/2$ for different RDDI strengths, (a) $\gamma_1=\gamma_2=1.5\beta$, (b) $\gamma_1=\gamma_2=6\beta$, (c) $\gamma_1=\gamma_2=10\beta$.}
\label{fig.5}
\end{figure*}

 For orthogonal dipole moments, the excited states of both atoms hybridize into two dressed states within PBG and the single state in the pass band  does not contribute to population oscillations. Notably, the probability amplitudes $A_{2,4}(t)$ [Eqs.(\ref{A.10}), (\ref{A.12}) in Appendix] (corresponding to $\ket{a_2,a_6}$ and $\ket{a_3,a_5}$, respectively) entirely vanish, suppressing transitions to the second excited state and restricting dynamics to $\ket{a_1} \rightarrow \ket{a_3} $$(\ket{a_4}\rightarrow \ket{a_6})$ subspace.
 For weaker RDDI rates ($\gamma_1=\gamma_2=1.5\beta$), the dressed states lie at $-1.5\beta\iota$, and $-1.1\beta\iota$ such that the atoms decouple almost immediately due to the effect of band edge DOS and entanglement vanishes faster as shown in Fig. \ref{fig.5}(a). For a stronger dipole-dipole coupling ($\gamma_1=\gamma_2=6\beta$), the localized states in the band gap lie at $-6\beta\iota$ and $-5.6\beta\iota$ and for $\gamma_1=\gamma_2=10\beta$, the dressed states lie at $-10\beta\iota$ and $-9.6\beta\iota$, deeper in the band gap region [see Fig. \ref{fig.6}(a)]. We observe that the oscillatory behaviour for $\gamma_1=\gamma_2=10\beta$ persists up to a normalized interaction timescale of $\beta t \approx 2000$ and then undergoes exponential damping [see Fig. \ref{fig.5}(c)]. This behaviour reveals a quantum state redistribution process where atom-photon correlations mediate transient interatomic entanglement generation through RDDI. This interplay facilitates coherent energy exchange while dissipative coupling to leaky evanescent modes enforces eventual decoherence. Hence, the strength of RDDI allows sustained radiative exchange and continued resonant interaction that preserves entanglement for longer times.
 To the best of our knowledge, this represents the first demonstration of such non-Markovian, reservoir-engineered entanglement dynamics in such cavity QED systems leveraging multi-level atoms. The measured lifetime of entanglement approaches the timescale recently demonstrated in giant atom platforms, where extended dipole moments and waveguide-mediated interactions enhance entanglement preservation \cite{PhysRevLett.130.053601,PhysRevResearch.5.043135}.

\begin{figure*}[!htbp]
\centering
\includegraphics[width=0.75\linewidth]{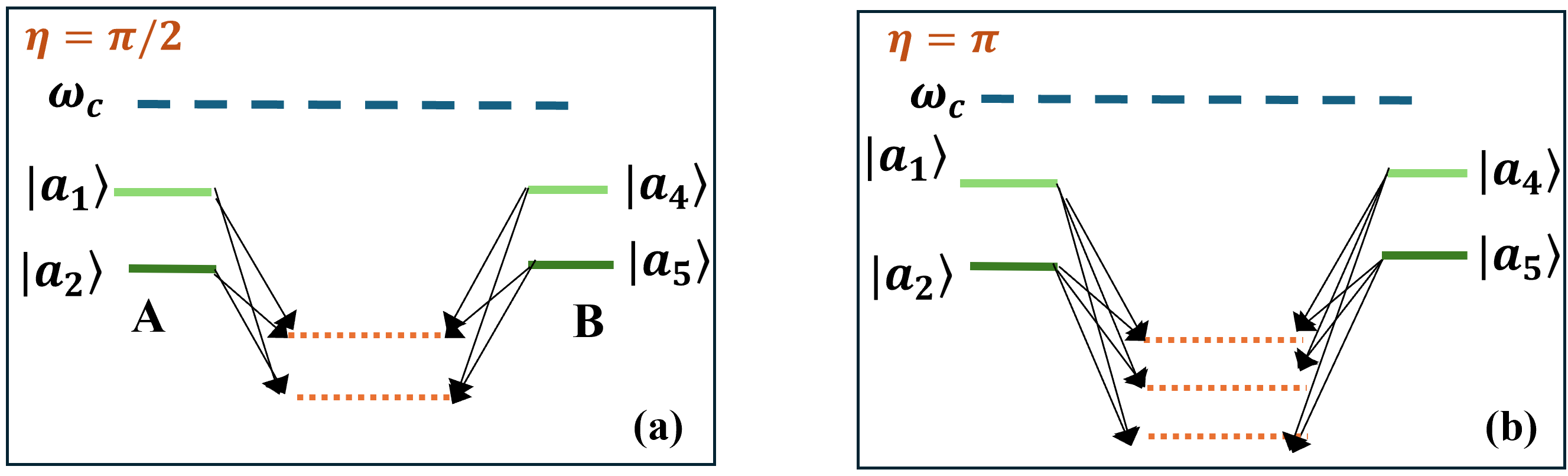}
\caption{Schematic of atomic splitting for the entangled bright state as initial state for the position of $\omega_{1c}=-0.6\beta$, $\omega_{2c}=-1\beta$ relative to $\omega_c$, where $\omega_c$ lie above $\ket{a_1}(\ket{a_4})$ and $\ket{a_2}(\ket{a_5})$, and $\omega_{12}=0.4\beta$, when dipole transitions of excited levels $\ket{a_1}(\ket{a_4})$ and $\ket{a_2}(\ket{a_5})$ are (a) orthogonal, $\eta=\pi/2$, (b) anti-parallel, $\eta=0$.} 
\label{fig.6}
\end{figure*}


We also emphasize that, for the orthogonal dipole arrangement with an initially entangled bright state, the atom-photon bound states formed show strong spatial localization of photons and share features analogous to dark states discussed in \cite{PhysRevX.12.011054,PhysRevLett.134.133603}. However, it is important to note that despite these similarities, their underlying physical origins differ.
Notably, RDDI does not merely stabilize these states; it also induces coherent oscillations of the trapped population which results from the interplay between atomic excitations and the spatially localized photon field. The spatial confinement provided by PBG environment strongly suppresses spontaneous emission, thereby offering a promising platform for robust storage of quantum excitations. However, realizing and harnessing these effects in practice will require further detailed investigation \cite{PhysRevLett.129.253601}.
\begin{figure*}[!htbp]
\centering
\includegraphics[width=0.7\linewidth]{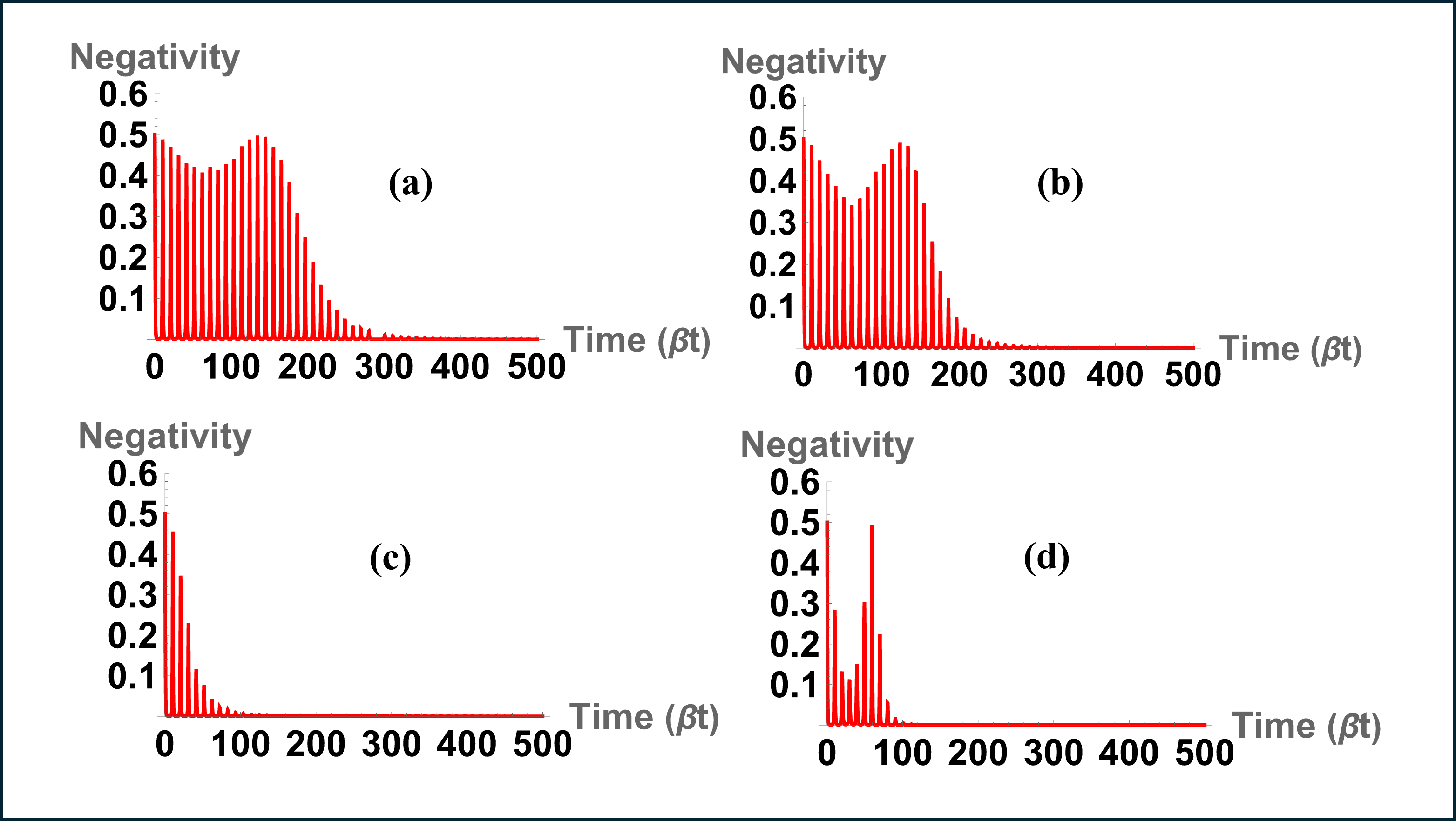}
\caption{Negativity dynamics as a function of $\beta t$ for different positions of upper levels, $\omega_{1c}$ and $\omega_{2c}$ relative to band edge $\omega_c$ for orthogonal dipole transitions, (a) $\omega_{1c}=0.6\beta$, $\omega_{2c}=0.2\beta$, (b) $\omega_{1c}=0.6\beta$, $\omega_{2c}=-0.4\beta$, (c) $\omega_{1c}=-0.6\beta$, $\omega_{2c}=-1\beta$, (d) $\omega_{1c}=-1.6\beta$, $\omega_{2c}=-2.6\beta$. The strength of RDDI considered is $\gamma_1=\gamma_2=5\beta$.}
\label{fig.7}
\end{figure*}

\subsection{Effect of positioning of excited levels relative to band edge:}
We further explore the effect of positioning of atomic excited states $\omega_{1c}$ and $\omega_{2c}$ relative to band edge $\omega_c$ for entangled bright initial state and $\eta=\pi/2$ configuration. We observe that the position of excited states critically governs entanglement dynamics through the influence of RDDI and photonic DOS. As shown in Fig. \ref{fig.7}(c), when both transitions lie within the deeper band gap region, the RDDI-mediated entanglement decays rapidly compared to configurations where transitions reside in the band pass region [Fig. \ref{fig.7}(a)] or straddle the band edge [Fig. \ref{fig.7}(b)]. This accelerated decay arises from the exponential suppression of the photonic DOS within the band gap, which inhibits photon-mediated coherent coupling while permitting residual dissipative processes via evanescent modes. Similar result has been reported for two qubits in \cite{PhysRevA.89.062117}. Further, preservation of a finite amount of entanglement has also been reported in band gap for two NV (qubits) centers in diamond nano-crystal \cite{PhysRevA.87.022312}.
Notably, deeper placement of excited states $\omega_{1c}=-1.6\beta$ and $\omega_{2c}=-2.6\beta$ induces a transient entanglement revival due to hybridized light-matter states localized near atomic positions. However, the overall decay rate remains faster than in shallower band gap configurations, reflecting a trade-off between suppressed radiative decay and diminished RDDI efficiency. The observed faster entanglement loss with increasing band gap depth underscores the competition between two mechanisms: Localized mode coupling in which evanescent fields mediate short-range RDDI, but their penetration depth decreases exponentially with depth in the band gap \cite{doi:10.1073/pnas.1603788113}; non-Markovian feedback in which band-edge proximity enables delayed photon reabsorption, while deeper band gap placement quenches this effect \cite{liu2017quantum}. This primarily occurs because the lack of resonant normal modes prevents atoms from exchanging real photons, so that energy transfer is mediated instead by very slow oscillations of virtual photons associated with non-resonant modes \cite{PhysRevA.42.2915}. Therefore, RDDI plays a crucial role near the band edge DOS, while its influence diminishes as one moves deeper into the bandgap region. In addition, these results also highlight the delicate balance between coherent interactions and dissipative processes in structured photonic environments.

\section{Conclusion}\label{sec:4}
This study elucidates the synergistic dynamics between RDDI and quantum interference in multi-level atomic systems coupled to PCs, contrasting entanglement evolution from initially correlated versus separable states. Our analytical and numerical analysis reveals that RDDI and quantum interference serve as critical control parameters for tailoring entanglement dynamics in such structured photonic environments. Specifically, RDDI dominates when interatomic distances lie within the localization length of photonic bound states induced by the band gap, enabling evanescent-field-mediated coherent coupling. Conversely, quantum interference emerges as a key modulator when orthogonal dipole transitions interact with the engineered photonic DOS. We demonstrate that antiparallel dipole configurations universally enhance entanglement lifetimes for both separable and entangled initial states. This study identifies a newly characterized oscillatory entanglement generation mechanism in systems initialized with entangled states under orthogonal dipole alignment, where destructive interference between vacuum-mediated pathways generates periodic population transfer between hybridized dressed states. This phenomenon exhibits non-Markovian features such as delayed feedback and entanglement oscillations. In contrast, systems initialized with separable states exhibit accelerated decay of logarithmic negativity for orthogonal dipole configurations due to unmitigated radiative leakage through non-interfering decoherence channels. This dichotomy highlights the critical role of initial entanglement in activating interference-protected dissipation pathways. These insights are critical for designing PC-based quantum devices, where initial state preparation and dipole alignment dictate entanglement resilience. These findings highlight the potential of PC cavities as platforms for entanglement engineering, with applications including quantum switches, quantum clocks, quantum memories, and quantum sensors. 

These findings can be further explored for an asymmetry in PC, which generates a delta-resonance mode within the band gap. It is also possible to explore the influence of asymmetric RDDI, $\gamma_1 \neq \gamma_2$ on coherence transfer and the formation of dressed states.
There is also potential to investigate alternative initial superpositions of atomic states in multi-photon manifolds that could enhance the duration of entanglement, surpassing the traditional bright and dark state frameworks typically considered in cavity QED. 

Recent advancements have demonstrated both entanglement transport using neutral atoms with PC cavities \cite{doi:10.1126/science.abi9917} and high-fidelity, error-detected operations via optical cavities \cite{doi:10.1126/science.adr7075}, showcasing the strength of cavity-based systems for quantum information processing. While nanophotonic interfaces enable long-term entanglement during atom transport, optical cavities offer fast and accurate operations in stationary setups, making both approaches vital for future quantum technologies.

\appendix* 
\onecolumngrid
\section*{Appendix: Calculation of probability amplitudes $A_i(t)'s$ and $B_k(t)$}\label{Appendix}
\setcounter{equation}{0}
\renewcommand{\theequation}{A.\arabic{equation}}

Using Laplace transform in Eq. [\ref{eq:2}-\ref{eq:6}] we can obtain the following linear algebraic equations for $A_i(x)'s$ \cite{PhysRevA.61.043809}:
\begin{eqnarray}
    (x+\Gamma_{11})A_1(x) +  \Gamma_{12}A_2(x')+(\iota\gamma_1+\Gamma_{11})A_3(x) +\Gamma_{12}A_4(x') - A_1(0) &=& 0\;, \nonumber\\
   \Gamma_{12}A_1(x)+ (x'+\Gamma_{22})A_2(x')+ \Gamma_{12}A_3(x) + (\iota\gamma_2+\Gamma_{22})A_4(x')  - A_2(0) &=& 0 \;,\nonumber\\
    (\iota\gamma_1+\Gamma_{11})A_1(x) +\Gamma_{12}A_2(x')+ (x+\Gamma_{11})A_3(x)  + \Gamma_{12}A_4(x') - A_3(0)& =& 0 \;,\nonumber\\
   \Gamma_{12}A_1(x)  + (\iota\gamma_2+\Gamma_{22})A_2(x') + \Gamma_{12}A_3(x) +(x' +\Gamma_{22})A_4(x') - A_4(0) &=& 0 \;, \label{eqns}
\end{eqnarray} 
where $x'=x-\iota\omega_{12}$.
\begin{figure}[!htbp]
\centering
\includegraphics[width=0.25\linewidth]{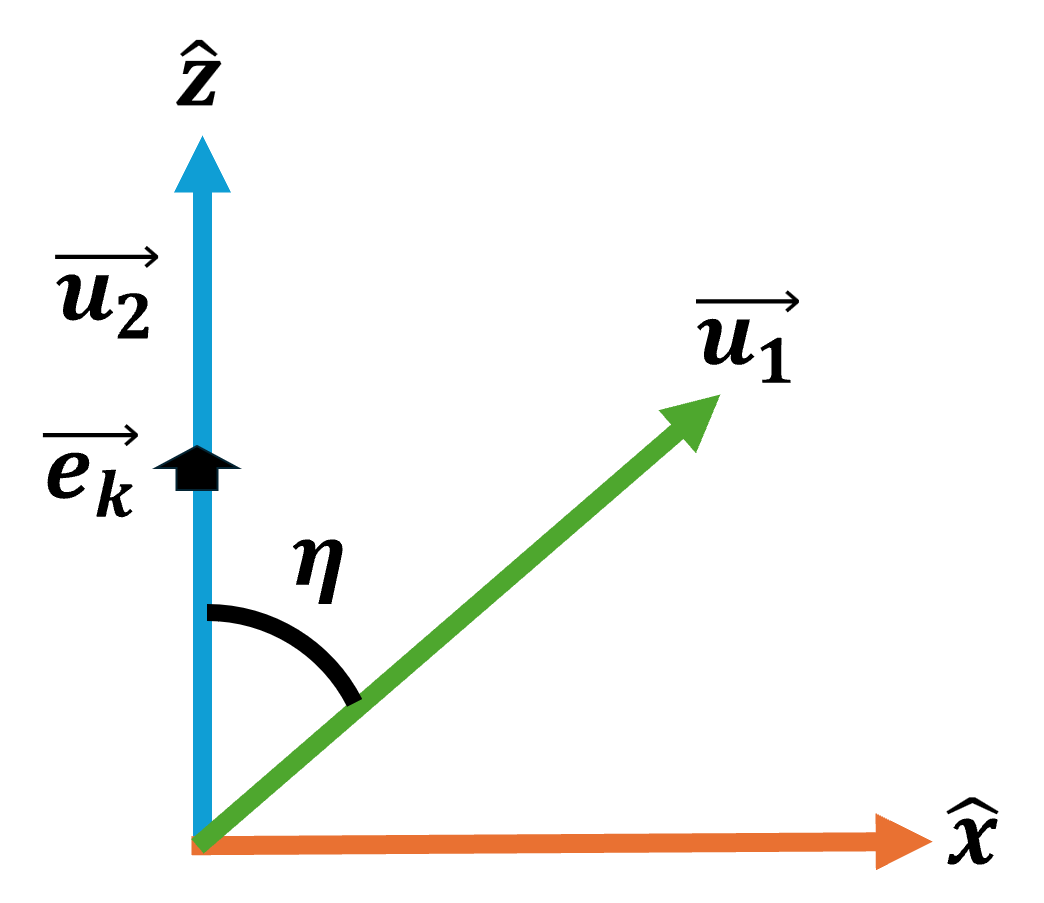}
\caption{The geometry of two dipole moment transitions of each atom showing the direction of $\overrightarrow{u_1}$ and $\overrightarrow{u_2}$ with respect to axes.}
\label{fig.A1}
\end{figure}
Here $\Gamma_{ij}=\sum_k\frac{g_k^{(i)} g_k^{(j)}}{[x-\iota(\omega_{13}-\omega_k)]}$, $(i,j=1,2)$ is obtained after performing Laplace transform. Using dispersion relation, converting summation over transverse plane waves into an integral and performing the integral \cite{PhysRevA.61.043809}, we obtain
\begin{equation}
\Gamma_{ii} =\frac{\beta_i^\frac{3}{2}}{\iota \sqrt{-\iota x -\omega_{1c}}}\;\;;\;\;
\Gamma_{ij} = \frac{(\beta_i \beta_j)^\frac{3}{4}}{\iota \sqrt{-\iota x -\omega_{1c}}} (\overrightarrow{e_k}.\overrightarrow{u_1}).(\overrightarrow{e_k}.\overrightarrow{u_2})\;, i\neq j\;.\label{gammaij}
\end{equation} 
Here, $\overrightarrow{u_j}$ are unit vectors of atomic dipole moments. We choose $\overrightarrow{u_1} = (0,0,1)$, $\overrightarrow{u_2} = (\sin{\eta},0,\cos{\eta})$  without any loss of generalization. The $\overrightarrow{e_k}$ is the unit vector along the propagation direction $\vec{k}=(k\sin{\theta}\cos{\phi},k\sin{\theta}\sin{\phi},k\cos{\theta})$ as shown in Fig. \ref{fig.A1}. This implies $\overrightarrow{e_k}.\overrightarrow{u_2} = \cos{\eta}$, $\omega_{12}=\omega_{13}-\omega_{23}$, $\omega_{1c}=\omega_{13}-\omega_c$, $\omega_{2c}=\omega_{23}-\omega_c$, and 
\begin{equation}\label{beta}
\beta_i^{3/2} = \frac{(\omega_i d_i)^2}{6 \pi\epsilon_0\hbar} \frac{k_0^3}{\omega_c^{3/2}}. 
\end{equation}\\
\noindent
Solving the equations (\ref{eqns}), we get $A_i(x)'s$ as
\begin{eqnarray}
    DA_1(x)& =& \Gamma_{12}(x-\iota\gamma_1)\left[A_2(0)+A_4(0)\right] - 2\Gamma_{12}^2\left[A_1(0)+A_3(0)\right]\nonumber\\
    &&- A_3(0)(\iota\gamma_1+\Gamma_{11})(x'+\iota\gamma_1+2\Gamma_{22}) + A_1(0)(x+\Gamma_{11})(x'+\iota\gamma_2+2\Gamma_{22})\;, \label{A.1} \\
    DA_2(x') &=& 2\Gamma_{12}^2\left[A_2(0)+A_4(0)\right]-\Gamma_{22}(x'-\iota\gamma_2)\left[A_1(0)+A_3(0)\right]\nonumber\\
    &&+(x+\iota\gamma_1+2\Gamma_{11})\left[A_2(0)(x'+\Gamma_{22})-A_4(0)(\iota\gamma_2+\Gamma_{22})\right] \;,\label{A.2}\\
   D A_3(x)& =& \Gamma_{12}(x-\iota\gamma_1)\left[A_2(0)+A_4(0)\right] - 2\Gamma_{12}^2\left[A_1(0)+A_3(0)\right]\nonumber\\
    &&- A_1(0)(\iota\gamma_1+\Gamma_{11})(x'+\iota\gamma_1+2\Gamma_{22}) + A_3(0)(x+\Gamma_{11})(x'+\iota\gamma_2+2\Gamma_{22})  \;,\label{A.3}\\
   D A_4(x')& = &2\Gamma_{12}^2\left[A_2(0)+A_4(0)\right] -\Gamma_{22}(x'-\iota\gamma_2)\left[A_1(0)+A_3(0)\right]\nonumber\\
    &&-(x+\iota\gamma_1+2\Gamma_{11})\left[A_2(0)(\iota \gamma_2+\Gamma_{22})-A_4(0)(x'+\Gamma_{22})\right]\;,\label{A.4}
\end{eqnarray}
where $D=(x-\iota\gamma_1)\{(x+\iota\gamma_1+2\Gamma_{11})(x'+\iota\gamma_2+2\Gamma_{22})-4\Gamma_{12}^2)\} $.

Using inverse Laplace transform, the probability amplitudes can then be obtained as 
\begin{eqnarray}
A_{1,3}(t)& =& \frac{1}{2\pi\iota} \int_{\sigma-\iota\infty}^{\sigma+\iota\infty} A_{1,3}(x) e^{xt} \, dx \;,\label{A.7}\\
A_{2,4}(t) &=& \frac{1}{2\pi\iota} \int_{\sigma-\iota\infty}^{\sigma+\iota\infty} A_{2,4}(x-\iota\omega_{12}) e^{xt} \, dx \;,\label{A.8}
\end{eqnarray}
where $\sigma$ is a real constant that exceeds the real part of all the singularities of $A_i(x)'s$. 

Solving contour integration in Eqs. (\ref{A.7}) and (\ref{A.8}) \cite{PhysRevA.61.043809}, we get $A_i(t)'s$ in the following forms:
\begin{eqnarray}
A_1(t) &=& \sum_{j} \frac{f_1\left(x_{G_j}^{(1)}\right)}{G_1'\left(x_{G_j}^{(1)}\right)} e^{x_{G_j}^{(1)}t} + \sum_j \frac{f_2(x_{G_j}^{(2)})}{G_2'(x_{G_j}^{(2)})} e^{x_{G_j}^{(2)}t} \nonumber\\
&&+ \frac{1}{2\pi\iota} e^{\iota\omega_{1c}t} \int_{0}^{\infty} \left(\frac{f_1(-x+\iota\omega_{1c})}{G_1(-x+\iota\omega_{1c})}-\frac{f_2(-x+\iota\omega_{1c})}{G_2(-x+\iota\omega_{1c})}\right)e^{-xt} \, dx  \;,\label{A.10} \\
A_2(t) &=& \sum_l \frac{f_3(x_{H_j}^{(1)})}{H_1'(x_{H_j}^{(1)})} e^{x_{H_j}^{(1)}t} + \sum_l \frac{f_4(x_{H_j}^{(2)})}{H_2'(x_{H_j}^{(2)})} e^{x_{H_j}^{(2)}t} \nonumber\\
&&+ \frac{1}{2\pi\iota} e^{\iota\omega_{1c}t} \int_{0}^{\infty} \left(\frac{f_3(-x+\iota\omega_{1c})}{H_1(-x+\iota\omega_{1c})}-\frac{f_4(-x+\iota\omega_{1c})}{H_2(-x+\iota\omega_{1c})}\right)e^{-xt} \, dx \;,\label{A.11} \\
A_3(t) &=& \sum_j \frac{f_5(x_{G_j}^{(1)})}{G_1'(x_{G_j}^{(1)})} e^{x_{G_j}^{(1)}t} + \sum_j \frac{f_5(x_{G_j}^{(2)})}{G_2'(x_{G_j}^{(2)})} e^{x_{G_j}^{(2)}t} \nonumber\\
&&+ \frac{1}{2\pi\iota} e^{\iota\omega_{1c}t} \int_{0}^{\infty} \left(\frac{f_5(-x+\iota\omega_{1c})}{G_1(-x+\iota\omega_{1c})}-\frac{f_6(-x+\iota\omega_{1c})}{G_2(-x+\iota\omega_{1c})}\right)e^{-xt} \, dx \;,\label{A.12} \\
A_4(t) &=& \sum_l \frac{f_7(x_{H_j}^{(1)})}{H_1'(x_{H_j}^{(1)})} e^{x_{H_j}^{(1)}t} + \sum_l \frac{f_8(x_{H_j}^{(2)})}{H_2'(x_{H_j}^{(2)})} e^{x_{H_j}^{(2)}t} \nonumber\\
&&+ \frac{1}{2\pi\iota} e^{\iota\omega_{1c}t} \int_{0}^{\infty} \left(\frac{f_7(-x+\iota\omega_{1c})}{H_1(-x+\iota\omega_{1c})}-\frac{f_8(-x+\iota\omega_{1c})}{H_2(-x+\iota\omega_{1c})}\right)e^{-xt} \, dx \;,\label{A.13} 
\end{eqnarray}
\\
Similarly, $B_k(t)$ can be calculated by directly integrating eq.(\ref{eq:6}) after substitution of $A_1(t),A_2(t),A_3(t)$, and $A_4(t)$ from Eqs. (\ref{A.10} - \ref{A.13}):
\begin{equation}\label{B_k(t)}
 B_k(t) = g_k^{(1)}\int_{0}^{\infty}\left\{A_1(t)+ A_3(t)\right\} e^{-\iota (\omega_{13}-\omega_{k})t}  dt + g_k^{(2)}\int_{0}^{\infty}\left\{A_2(t)  + A_4(t) \right\}e^{-\iota (\omega_{23}-\omega_{k})t} dt
\end{equation}\\
Here, $\beta'=\frac{\beta^{3/2}}{\iota\sqrt{-\iota x-\omega_{1c}}}$ (where $\beta_1=\beta_2=\beta$). The relevant functions in Eqs. (\ref{A.10}-\ref{A.13}) are given by \\
\begin{eqnarray}
f_{1,2}(x) &=& -\iota\left(\iota x+\omega_{12}-\gamma_2+2\iota\beta'\right)\left[\left(x+\beta'\right)A_1(0)-\iota\left(\gamma_1-\iota\beta'\right)A_3(0)\right] \nonumber\\
&& \pm \beta'\cos{\eta}\left[2\beta\cos{\eta}\left(A_3(0)-A_1(0)\right) \pm \iota\left(\iota x+\gamma_1\right)\left(A_2(0)+A_4(0)\right)\right] \;,\nonumber \\
f_{3,4}(x) &=& -\iota\left(\iota x-\gamma_1+2\iota\beta'\right)\left[\left(x+\beta'\right)A_2(0)-\iota\left(\gamma_2-\iota\beta'\right)A_4(0)\right] \nonumber\\
&& \pm \beta'\cos{\eta}\left[2\beta\cos{\eta}\left(A_2(0)-A_4(0)\right) \pm \iota\left(\iota x+\omega_{12}+\gamma_2\right)\left(A_3(0)+A_1(0)\right)\right] \;, \nonumber\\
f_{5,6}(x) &=& -\iota\left(\iota x+\omega_{12}-\gamma_2+2\iota\beta'\right)\left[\left(x+\beta'\right)A_3(0)-\iota\left(\gamma_1-\iota\beta'\right)A_1(0)\right] \nonumber\\
&& \pm \beta'\cos{\eta}\left[2\beta\cos{\eta}\left(A_1(0)-A_3(0)\right) \pm \iota\left(\iota x+\gamma_1\right)\left(A_2(0)+A_4(0)\right)\right] \;,\nonumber \\
f_{7,8}(x) &=& -\iota\left(\iota x-\gamma_1+2\iota\beta'\right)\left[\left(x+\beta'\right)A_4(0)-\iota\left(\gamma_2-\iota\beta'\right)A_2(0)\right] \nonumber\\
&& \pm \beta'\cos{\eta}\left[2\beta\cos{\eta}\left(A_4(0)-A_2(0)\right) \pm \iota\left(\iota x+\omega_{12}+\gamma_2\right)\left(A_1(0)+A_3(0)\right)\right] \;, \nonumber 
\end{eqnarray}
\begin{eqnarray}
G_{1,2}(x) &=& \pm \iota\left(\iota x+\gamma_1\right)\left[\left(\iota x-\gamma_1\right)\left(\iota x+\omega_{12}-\gamma_2\right)\left(1 \pm 2\beta'\right)-4\beta'^2\sin^2{\eta}\right] \;,\label{A.5} \\
H_{1,2}(x) &=& \pm \iota\left(\iota x+\omega_{12}+\gamma_2\right)\left[\left(\iota x-\gamma_1\right)\left(\iota x+\omega_{12}-\gamma_2\right)\left(1 \pm 2\beta'\right)-4\beta'^2\sin^2{\eta}\right] \;.\label{A.6}
\end{eqnarray}

The poles of $G_{1,2}(x)$ and $H_{1,2}(x)$ are denoted as $x_{G_j}^{(1,2)}$ and $x_{H_j}^{(1,2)}$ respectively and can be classified into localized roots (pure imaginary roots with imaginary part larger than $\omega_{1c}$) and propagating roots (complex roots with negative real part and imaginary part smaller than $\omega_{1c}$). The number of roots depends on the position of atomic upper levels relative to band edge ($\omega_{1c}$ and $\omega_{2c}$) which can be either $\omega_{1c} \geq \omega_{12}/2$ or $\omega_{1c} < \omega_{12}/2$ as described in \cite{PhysRevA.61.043809}.

Note that the expressions for $A_i(t)$ contain the contribution of all existing poles of the functions $G_{1,2}(x)=0$ and $H_{1,2}(x)=0$ ($x$ is the inverse variable to the time $t$, used while doing Laplace transformations), and the integral along the real axis of $x$ explicitly encodes the decay rate information. 

\twocolumngrid

\bibliographystyle{apsrev4-2}
\bibliography{references}

\end{document}